\documentclass[rnote]{aa}
\usepackage{graphicx}
\usepackage{txfonts}

\begin{document}

\title{Direct imaging of the young spectroscopic binary HD\,160934
\thanks{Based on observations collected at the Centro Astron\'omico Hispano Alem\'an (CAHA) at
	Calar Alto, operated jointly by the Max-Planck-Institut f\"ur Astronomie and the Instituto
	de Astrof\'isica de Andaluc\'ia (CSIC). Based on observations obtained 
             with MPIA's K\"onigstuhl mountain 70cm telescope.}}
          
\author{Felix Hormuth\inst{1} \and
          	Wolfgang Brandner\inst{1,2} \and
          	Stefan Hippler\inst{1} \and
          	Markus Janson\inst{1}\and
		Thomas Henning\inst{1}
          	}
\offprints{Felix Hormuth}

\institute{Max-Planck-Institut f\"ur Astronomie, K\"onigstuhl 17,
         		69117 Heidelberg, Germany\\
         		\email{hormuth@mpia.de}
		\and
		UCLA, Division of Astronomy, Los Angeles, CA 90095-1547, USA}

\date{Received ---; accepted ---}

\abstract
{Members of nearby moving groups are promising candidates for the detection
 of stellar or substellar components by direct imaging. Mass estimates and
 magnitude measurements of detected companions to young stars are valuable 
 input data to facilitate the refinement of existing pre-main-sequence stellar models. In this
 paper we report on our detection of a close companion to \object{HD\,160934},
 a young active star, SB1 spectroscopic binary and a suggested member of the AB Doradus
 moving group.}
{We obtained high angular resolution images of nearby young stars, searching for close
 companions. In the case of HD\,160934, direct imaging is combined with unresolved
 photometry to derive mass estimates.}
{High angular resolution was achieved by means of the so-called ``Lucky Imaging''
technique, allowing direct imaging close to the diffraction limit in the SDSS z' band.
Our results are combined with pre-discovery HST archive data, own UBV(RI)$_C$ broadband
photometry, published JHK magnitudes and available radial velocity measurements to
constrain the physical properties of the HD\,160934 close binary.}
{At an assumed age of $\sim$80\,Myr, we derive mass estimates of 0.69\,M$_{\sun}$ and
0.57\,M$_{\sun}$, respectively, for HD\,160934 and its close companion. We suggest that
the direct detection may be identical to the spectroscopically discovered companion,
leading to a period estimate of $\sim$8.5~years and a semimajor axis of a$\approx$4.5\,A.U.}
{}


\maketitle

\section{Introduction}

HD\,160934 (=\,\object{HIP\,86346}) is a young late-type star with a spectral type of K7 to M0
(Reid et al.\ \cite{reid95}; Zuckerman et al.\ \cite{zuckerman04a}) at
a distance of $\approx$24.5\,pc (Perryman et al.\ \cite{perryman97}). It
is chromospherically active (Mulliss \& Bopp \cite{mulliss94}) with
prominent EUV (e.g., Pounds et al.\ \cite{pounds93}) as well as X-ray emission
with an X-ray luminosity of L$_{\rm X}$ = $3.4 \times 10^{22}$\,W (H\"unsch
et al.\ \cite{huensch99}). The activity can also be traced in the 
H$\alpha$ line, which is seen in emission with an equivalent width between 
-0.09 and -0.13\,nm \ (Mulliss \& Bopp \cite{mulliss94}; 
Gizis et al.\ \cite{gizis02}; Zuckerman et al.\ \cite{zuckerman04a}). 
The detection of the Li~6708\AA \ line with an  equivalent width of 40\,m\AA \
(Zuckerman et al.\ \cite{zuckerman04a}) gives further evidence that HD\,160934 is
a relatively young star. These youth indicators combined with the 3d space
motion led Zuckerman et al.\ (\cite{zuckerman04a,zuckerman04b}) and Lopez-Santiago et al.\ 
(\cite{lopez06}) to suggest that HD\,160934 might be a member of the 
$\approx$50\,Myr old AB\,Dor moving group.

Because of its proximity to the Sun and its young age, HD\,160934 is
a good candidate for the direct detection of substellar, or even planetary 
mass companions. McCarthy \& Zuckerman (\cite{mccarthy04}) report that no
brown dwarf companion could be found at projected separations larger than 
75\,A.U.\ as a result of a near infrared coronagraphic study 
carried out at the Keck observatory. Using HST/NICMOS in coronagraphic mode,
Lowrance et al.\ (\cite{lowrance05}) report the detection of a possible wide
companion to HD\,160934 at a projected separation of $\approx$8\farcs7
(corresponding to $\approx$210\,A.U.) and at a position angle 
of $\approx$235\degr.
The brightness difference between the companion candidate, designated 
HD\,160934\,B, and HD\,160934\,A is $\Delta$H = 9.2\,mag. Under the assumption
that HD\,160934\,B constitutes a physical companion to  HD\,160934, 
Lowrance et al.\ (\cite{lowrance05}) derive a mass estimate of 
$\approx$0.15\,M$_{\sun}$ for this companion.

By combining 37 radial velocity measurements, G\'alvez et al. (\cite{galvez06}) 
were able to identify HD160934 as a spectroscopic SB1 binary and suggested
a period of P=6246.2318 days, a high eccentricity of e=0.8028, and a spectral
type of M2-M3V for the close companion, so that HD~160934 may be actually
at least a triple system.

In July 2006, we started a high angular resolution survey for close stellar and 
substellar companions to young nearby stars, taking advantage of the 
diffraction limited performance facilitated by using `Lucky Imaging' at the 
Calar Alto~2.2\,m telescope. The absence of a coronagraphic mask in our
set-up, and the fact that the diffraction limit of a 2.2\,m telescope
at the effective observation wavelength of 900\,nm corresponds to 
an angular resolution of $\approx$0\farcs1
means that our survey is sensitive to companions at close separations. 
In this {\it Research Note} we report on the first results of the survey, namely the 
direct imaging of HD\,160934 and a close companion. We summarise the
available photometric measurements, give new UBV(RI)$_C$z' magnitudes,
and discuss the possibility that the spectroscopic and our direct detection refer
to the same companion.

\section{Observations and data reduction}

\begin{table}
\caption{Direct imaging observing log for HD 160934.}
\label{tbl:obslog}
\centering
\begin{tabular}{ l c c c }
\hline\hline
\noalign{\smallskip}
Date &Telesc./Inst. & Filter & t$_{\rm int}$  \\ 
\noalign{\smallskip}
\hline
\noalign{\smallskip}
June, 30 1998       &  HST/NIC2 & F165M  & $2 \times 0.626$s \\
July, 8 2006        &   CA 2.2m/AstraLux & RG780  & 4.5s$^a$ \\
July, 8 2006        &   CA 2.2m/AstraLux & RG830  & 4.5s$^a$ \\
\noalign{\smallskip}
\hline
\end{tabular}
\begin{list}{}{}
\item[$^{\mathrm{a}}$]
2\% best of 15000 frames with an individual t$_{\rm exp} = 0.015$s
\end{list}
\end{table}

\begin{figure*}[htb]
\centering
 \includegraphics[height=18.0cm,angle=90]{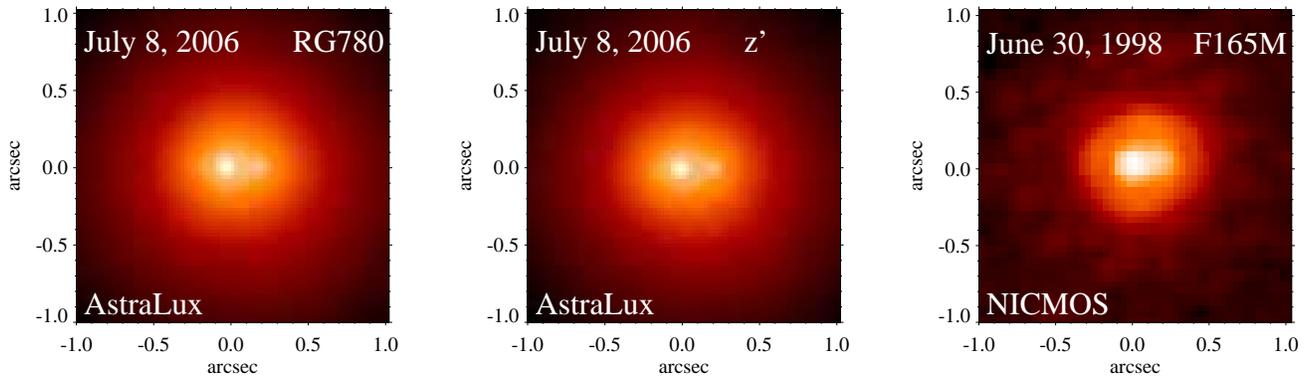}
\caption{AstraLux RG780, z' and NICMOS F165M images of the binary using a logarithmic
intensity scale. The field of view is $2'' \times 2''$, and North is up and East
 is to the left. The binary components are clearly detectable in all three 
bands. Note that both the AstraLux and NICMOS images have been oversampled by 
a factor of 2 compared to the original instrumental pixel scale as part of the 
drizzle process.}
\label{fig1}
\end{figure*}

\subsection{Direct imaging with AstraLux at the Calar Alto 2.2m telescope}

Observations at the Calar Alto 2.2m telescope were obtained with
the new 'Lucky Imaging' instrument AstraLux. The Lucky Imaging
technique is based on the fact that in a large set of ground based short 
exposure images at least a small fraction will be relatively unaffected 
by atmospheric turbulence. By finding and using only these high-quality 
images, one is able to obtain diffraction limited images from the ground 
without the need of adaptive optics (Baldwin et al.\ \cite{baldwin01}; 
Tubbs et al.\ \cite{tubbs02}; Law et al.\ \cite{law06}).

The AstraLux instrument basically consists of an electron multiplying, thinned and 
back-illuminated CCD, which is operated in frame transfer mode and capable 
of running at a frame rate of 34\,Hz when reading the full 512x512 pixel frame. The
camera is mounted at the Cassegrain focus of the telescope behind
a simple Barlow lens assembly, which changes the focal ratio from f/8 to f/32
in order to get an image scale of 46.6\,mas/pixel at a physical pixel size of
16$\times$16$\mu$m$^2$. This provides
Nyquist sampling at $\lambda \approx 1\mu{\rm m}$, and slightly
undersamples the theoretical point spread function (PSF) at the typical observation
wavelengths of 700-950\,nm. With the exception of the mount needed to attach 
the camera to the telescope, all components were bought off-the-shelf, making
AstraLux a simple and cost-effective instrument, taking less than six months from the 
first idea to first light.

HD 160934 was observed as part of a larger sample of nearby young
stars on July 8, 2006 under good  atmospheric conditions with a 
photometric sky and a V-band seeing around 0\farcs8.
The raw data for HD 160934 consists of two FITS cubes of 15000 frames 
with 15ms exposure time each. We used only a 256x256 pixel subarray of the
chip, leading to a frame rate of $\approx$65\,Hz. The target
was observed through two different filters, namely Schott RG780 
and RG830 longpass filters, the latter roughly matching the SDSS z'-band (for the
definition of the SDSS passbands, see e.g. Fukugita et~al. \cite{fukugita96}).

The best images were selected by calculating the ratio between peak flux and
total flux of the observed star, referred to a circular aperture centered on the
brightest pixel. This number is linearly related to the Strehl number -- at least
in the high-quality images we are interested in -- and a quite reliable measure to
find the frames least affected by atmospheric turbulence. \\
The final image was constructed by combining
the best 2\% of all images using the Drizzle algorithm (Fruchter \& Hook 
\cite{fruchter02}), corresponding to a total integration time of 4.5 sec. 
The images were two times oversampled during
drizzling, leading to a final pixel scale of 23.3 mas/px.
The total time needed for data acquisition amounts to about 5\,min per filter,
including instrumental overheads, and approximately the same time
is needed for data reduction. Our own version of the drizzle
algorithm makes use of the fact that we do not expect any changes
in image rotation or field distortion between the individual images,
and thus only considers translations. This allows a fast and simple implementation,
and a near real-time data reduction.

The pixel scale and image orientation were determined with a set of seven
'standard binary' observations, i.e., observations of known
visual binaries (chosen from the Sixth Catalog of Orbits of Visual Binary
Stars (Hartkopf \& Mason \cite{hartkopf06}), orbit grades 1 and 2) with
angular separations between 0\farcs6 and 2\farcs8. These binaries were
observed several times during the night (typically 2 standards/hour)
in the same manner and with the same instrumental setup as the 
science targets, and the final reference images were obtained
using the same reduction techniques.
The pixel scale and rotation angle determined using these reference
images showed a high stability within 1\% and 0.5\degr,
respectively, during the whole night. Although utilizing double stars as
calibrators for other double star measurements is far from optimal,
we think that using a set of several stars is sufficient to determine sound upper
limits of the measurement errors. 

\subsection{HST/NICMOS}

Observations of HD 160934 with HST/NICMOS were obtained on June 30, 1998 
(GO 7226, PI E.\ Becklin). Since the main aim of the programme was to search 
for faint, substellar companions, the coronagraphic mask in NIC2 was used. 
Because of this, the close companion to HD 160934 was not detected in the 
science data (see Lowrance et al.\ 2005). It is, however, detectable in the 
two acquisition frames. 

The acquisition frames were obtained with the F165M filter and an integration
time of 0.626\,s. For the analysis we assumed a pixel scale of 75.10\,mas/pixel.
The FWHM of the PSF was around 2 pixel, corresponding to 150\,mas, which is
close to the diffraction limit of HST at the observing wavelength. For the
second acquisition frame, taken at the beginning of a subsequent orbit, 
the HST guide star acquisition partially failed, resulting in
a slightly trailed PSF. This, however, did not strongly bias the analysis
of the data.

Figure \ref{fig1}
shows a comparison of the AstraLux RG\,780 and RG\,830, and the NICMOS F165M 
images of the HD 160934 binary, while Table~\ref{tbl:obslog} gives the
dates and details of the direct imaging observations. 

The analysis of the binary properties is based on the pipeline reduced frames.
Tiny Tim Version 6.3 (Krist \& Hook \cite{krist04}) was used to compute the 
theoretical PSF. In order to estimate the effect of HST ``breathing'' (i.e.\ 
focus changes induced by thermal expansion or shrinking of the optical train
of HST), also slightly defocussed PSFs were calculated and used for the binary
fitting. 

\subsection{Fitting of binary parameters}

For both data sets (AstraLux and NICMOS) a binary model (see Bouy et al.\ 
\cite{bouy03}) was fitted to the data 
in order to derive binary separation, position angle and brightness ratio.
For the NICMOS data, only the first acquisition frame was used. The slightly
trailed binary PSFs due to the partial guide star acquisition failure for the second
HST orbit resulted in a bias in the determination of the brightness ratio
using a non-trailed PSF. \\
While for the NICMOS data a theoretical model can serve as reference PSF
for binary fitting, this is not possible for the Lucky Imaging data. Since the PSF
shape depends strongly on actual seeing conditions, it is diffcult to predict
the theoretical PSF of a single star in the final results. Neither is there a 
single star available in our images, which
could have served as an accurate reference PSF. We therefore decided to use
a set of eight different reference PSFs, generated from observations of single
stars throughout the same night, and to use the weighted average of separation, brightness
ratio and position angle from all eight fit results, using the square of the residuals
as weighting parameter. While the residuals vary strongly for the different 
reference PSFs, the resulting separations and position angles show only a small
scatter. This is not quite the case for the measured brightness ratios - here the 1$\sigma$
error of the weighted mean is $\sim$16\%, considerably higher than for 
the HST-based F165M brightness ratio.
The derived fitting results can be found in Table \ref{binprop}. 

\begin{table}
\caption{Binary properties}
\label{binprop}
\centering
\begin{tabular}{ l c c c }
\hline\hline
\noalign{\smallskip}
Date & separation & PA    & brightn.\ ratio  \\
(Filter)     & [\arcsec] & [\degr] &   \\ 
\noalign{\smallskip}
\hline
\noalign{\smallskip}
June, 30 1998       &$0.155 \pm 0.001$   & $275.5 \pm 0.2$  & $0.485 \pm 0.006$  \\
(F165M)      &  & &  \\
July, 8 2006       &$0.215 \pm 0.002$   & $270.9 \pm 0.3$  &  $0.329 \pm 0.051$  \\
(RG830)       &  &  &  \\
\noalign{\smallskip}
\hline
\end{tabular}
\end{table}

\subsection{Unresolved photometry}
Available photometry of the unresolved binary covers the wavelength
range from U- to K-band. Weis (\cite{weis91}, \cite{weis93}) has published Johnson-Kron UBVRI
photometry. JHK magnitudes are contained in the 2MASS point source catalogue
(Skrutskie et al.~\cite{skrutskie06}). In addition, we obtained own UBV(RI)$_C$ and z'
photometry, using MPIA's K\"onigstuhl mountain 70cm telescope on 
September 5$^{th}$ and September 13$^{th}$ 2006. We also performed
narrowband photometry in H$\alpha$ in order to derive an H$\alpha$--R
color index, but calibration of this measurement proved to be rather
difficult and inaccurate, probably due to the relatively wide passband 
of the available H$\alpha$ interference filter.

During the course of the Hipparcos mission, 96 photometric measurements
of HD\,160934 were acquired. The lightcurve shows irregular brightness
variations with semi-amplitudes of 0.05-0.1\,mag on timescales on the
order of few days. Further evidence for the variability of HD\,160934 is given by 
Pandey et al. (\cite{pandey02}) and Henry et al. (\cite{henry95}), suggesting
amplitudes of $\approx$0.1mag. While Pandey et al. find a period of 43.2~days,
the observations of Henry et al. point to a much shorter value of 1.84~days with
some uncertainties regarding the presence of a longer period. \\
Table~\ref{tbl:phot} summarises all available unresolved photometric measurements.

\begin{table}
	\caption{Unresolved photometry of HD\,160934}
	\label{tbl:phot}
	\centering
	\begin{tabular}{l l r c c c}
	\hline\hline
	Source & Filter/Color & mag    & & 1$\sigma$  \\
	\hline
K\"onigstuhl 70cm & V          &   10.192 & $\pm$ & 0.014 \\
			     & U--B      &   0.947 &$\pm$& 0.008 \\
			     & B--V     &    1.215 &$\pm$& 0.005 \\
			     & V--R$_C$   &    0.789 &$\pm$& 0.004 \\
			     & R$_C$--I$_C$ & 0.766 &$\pm$&  0.007 \\
			     & z' & 8.820 &$\pm$& 0.009\\
\\
 Weis (\cite{weis91}, \cite{weis93})	&	V	&	10.28	&$\pm$& 0.020 \\
							&	U--B	&	0.95		&$\pm$& 0.015 \\	
							&	B--V   &   	1.23		&$\pm$& 0.015 \\
							&	V--R	&	0.78$^a$		&$\pm$& 0.015 \\
							&	R--I	&	0.63$^a$		&$\pm$& 0.015\\	
\\
2MASS		&	J	&	7.618	&$\pm$&	0.024 \\
			&	H	&	6.998	&$\pm$&	0.016 \\
			&	K	&	6.812	&$\pm$&	0.020 \\
\\
Hipparcos		&	V	&	10.29	&&\\
			&	B-V	&	1.591 &$\pm$&	0.400 \\
			&	V-I$_C$&	 2.58 &$\pm$&	0.91 \\	
\hline		
\end{tabular}
\begin{list}{}{}
\item[$^{\mathrm{a}}$]
The R and I-band photometry of Weis is given in the Kron system. Using the
cubic tranformations given by Bessell and Weis (\cite{bessell87}), the corresponding
colors in the Cousins system are V--R$_C$=0.78 and R$_C$--I$_C$=0.79.
\end{list}
\end{table}

\subsection{Radial velocity measurements}
\label{sec:radvel}
Radial velocity (RV) measurements of HD\,160934 exist for the years 1995--2004. 
G\'alvez et al. (\cite{galvez06}) published an RV curve based on 38 measurements, 
eventually leading to the classification of HD\,160934 as an SB1 spectroscopic binary
with an RV amplitude of K$\approx$7.2\,km\,s$^{-1}$.
They deduce a period of P=6246.2318~days and an eccentricity of e=0.8028, and derive
a spectral type of M2-M3V for the companion based on their mass estimates.
However, the observations cover less than one orbit and hence only 
one minimum of the radial velocity. In addition, the phase coverage is relatively sparse with 
only a single measurement for phases 0.2--0.9, and the given orbital parameters should
be considered as very preliminary values. While the available data certainly allows to
conclude that the orbit is relatively eccentric, it is this high eccentricity which makes period 
estimates without better coverage of the full RV curve unreliable.


\section{Physical properties of the HD 160934 binary}

\subsection{Common proper motion}
          
The proper motion of the HD 160934 main component amounts to 
$\mu_{\rm RA} = -31.25 \pm 14.43$mas/yr and 
$\mu_{\rm DEC} = 59.44 \pm 11.21$mas/yr. 
In the 8 years, which passed between the NICMOS and AstraLux observations, 
HD 160934 moved $250\pm 115$\,mas to the West, and $475\pm 90$\,mas to the
North. In the same period, the separation between HD 160934 A and c increased 
by $\approx 60$\,mas, and the position angle decreased by $\approx$5\,deg (see 
Table \ref{binprop}). This gives strong evidence that both sources form
indeed a physical binary.

\subsection{Photometric estimates of masses and spectral types}

Spectral types ranging from K7 (e.g., Reid et al.\ \cite{reid95}) to 
M0 (Zuckerman et al.\ \cite{zuckerman04a}) have been assigned to HD 160934.
While Reid et al.\ (\cite{reid95}) measured spectral indices,
Zuckerman  et al.\ (\cite{zuckerman04a}) based their spectral typing on the 
optical V--K colour of the unresolved binary. 

We derived our own estimates of the spectral types and components' masses
using the V--I$_C$ color index, the V magnitude of the unresolved system, and the SDSS z' and
F165M magnitude differences of the components. We compared these values
to the theoretical predicitions based on the models of Baraffe et al. (\cite{baraffe98}, abbreviated
BCAH98 in the following text) for
solar metallicity low-mass stars, searching for a mass combination best fitting
the available photometry. For our estimates we assumed coevality of the components.
Since the published BCAH98 models do not directly predict z'-band magnitudes, we use 
the empirical color transforms of Jordi et al. (\cite{jordi06}) to transform from R$_C$I$_C$
to SDSS z'. Though this seems like a crude method and calculating appropriate
z'-band magnitudes from model spectra should be preferred, we think that this
method is sufficient, since we are not fitting the z'-magnitudes directly, but the z'-magnitude difference
between the two components instead. For components of similar spectral types this
approximation is a valid way to convert between magnitude differences in the Johnson/Cousins
and the SDSS photometric system. 

A similar approach was taken in the case of the F165M magnitude differences. Since the
centers of the F165M and H passband are nearly identical, and the passbands differ only
in width, we directly compare the F165M magnitude difference with the H-band magnitude
differences of the models. As in the case of the z'-band difference, this is a sound
approximation for combinations of similar spectral types.

The BCAH98 model magnitudes were interpolated on a finer mass grid, and for each
possible mass combination in the range of 0.2-1.0M$_\odot$ the combined V magnitude,
V--I$_C$ color index, and the SDSS z' and H-band brightness ratios were computed and compared
to our measurements (in the case of the V-band magnitude, the comparison was made to
absolute V magnitude, based on the Hipparcos parallax).
The residuals were weighted by the measurement errors, and the best fitting mass
combination found by determining the global minimum of the residuals. We performed
fitting with models for different ages, namely in the range of 30 to 158~Myr. Minimum
residuals were obtained with the BCAH98 model for an age of 79~Myr, and the resulting
mass estimates  are given in Table \ref{tbl:massest}, together with the effective temperatures,
luminosities, and surface gravity  from BCAH98.

\begin{table}
\caption{HD160934Ac mass estimates for an assumed age of 79~Myr}
\label{tbl:massest}
\centering
\begin{tabular}{ c l l l l }
\hline\hline
\noalign{\smallskip}
Component   &    mass [M$_{\sun}$]    &      T$_{eff}$ [K] & log L/L$_{\sun}$ & g\\
\noalign{\smallskip}
\hline
\noalign{\smallskip}
A	&	0.69		&	4290 	&	-0.83		&	4.60 \\	
c	&	0.57		&	3780 	&	-1.23		&	4.68	\\
\noalign{\smallskip}
\hline
\end{tabular}
\end{table}

Using the modelled unresolved V-magnitude, we derive a distance module of M--m=2.81, 
corresponding to a distance of d=36.5pc or a parallax of $\pi=27.4$mas. Compared with the directly measured
Hipparcos parallax of 40.75$\pm$12.06mas, this deviates by 1.1$\sigma$. Of course, since
the Hipparcos parallax and its error were actually used in the fitting process, this photometric
distance estimate is not an independent measurement and somewhat circular. However,
the derived values constitute a set of physical parameters compatible to observations
within the measurement errors.

While it is tempting to believe that our data may be sufficient to find not only the components'
masses, but also to determine the age, it should be noted that especially the age is not very well
constrained by the data. Using the BCAH98 models for 50Myr results in nearly equally
small residuals, while the components' masses would then be  0.64 and 0.77~M$_{\sun}$,
respectively. The error of our mass estimates should therefore assumed to be in the
order of 0.1\,M$_{\sun}$. This clearly indicates that photometry-based mass estimates without
independent age constraints or spectroscopically determined effective temperatures are
affected by relatively high uncertainties. It should also be noted that the V magnitudes of the BCAH98
models are known to be be rather inaccurate for very-low-mass stars (e.g., Allard et al. \cite{allard97}), and that this may to some extent still be the case in the 0.7M$_{\sun}$ regime.

As a crosscheck, we computed the combined J, H and K magnitudes of the unresolved
binary as predicted by the BCAH98 models, and compare them to the 2MASS observations given 
in Table~\ref{tbl:phot}. The model magnitudes (using d=36.5pc) are J=7.59, H=6.99, K=6.86, which
gives a maximum deviation of 2.4$\sigma$ or 0.048mag in K-band. 

The derived values allow a tentative estimate of the components' spectral types, suggesting
a combination of a K5 and an M0 star, which is in good agreement with the published spectral types of
the unresolved binary (Reid et al. \cite{reid95}; Zuckerman et al. \cite{zuckerman04a}).

\subsection{Orbital parameters and comparison to RV data}
Between the HST and AstraLux observations, the change in projected separation and position
angle was 65mas and 4.6 degrees, respectively. This gives room for two possible scenarios: either
the orbital period is considerably larger than the 8 years time difference between the observations,
or it is an integer fraction of it (including $\approx 8$ years as one possibility). 
The spectroscopically determined period of P$\sim$17.1yr is -- at
first sight -- incompatible with this, since it would predict a difference in position angle of nearly
180 degrees for the direct imaging observations. A possible solution to this contradiction could
be that HD\,160934 is in fact a quadruple system (with the possible widely separated B component), 
and that the directly  imaged companion is not identical with the spectroscopic detection. 
However, as pointed out in Section~\ref{sec:radvel}, the available RV data does not allow
an unambiguous period determination. In fact, a period half the length of the suggested 17.1\,years
may still be compatible with the RV data and has to be confirmed or ruled out by further
observations.

If we assume a period of 8.55\,years, and use our mass estimates of M$_1$=0.69 and M$_2$=0.57\,M$_{\sun}$,
then the corresponding  semimajor axis would be a=4.5\,A.U. or 0\farcs12 at a distance of 36.5pc. Assuming
an eccentricity of e=0.8, this results in a maximum possible separation between the two components
of r=8.1\,A.U. or 0\farcs22, respectively. This is very close to the separation observed with AstraLux in
July 2006, which then means in return that the next periastron could be expected for around mid-2008.
Further RV measurements and resolved imaging in the next 2-3 years are necessary to sort
out the period ambiguity and to check whether the spectroscopically and directly detected
companions are indeed identical. 

\section{Conclusions}
By combining pre-discovery
HST archive data, our own high angular resolution astrometry and unresolved photometry,
we were able to derive mass and spectral type estimates for the HD\,160934 system. These
estimates are compatible with unresolved 2MASS photometry, Hipparcos
distance measurements, and existing age estimates for HD\,160934 and the AB Doradus
young moving group. We suggest that the directly imaged companion may be identical
with the companion discovered by radial velocity measurements, 
but with an orbital period of P$\approx$8.5\,years,
about half the value of the published period. Further high angular
resolution observations and radial velocity measurements in the next 2-3\,years will allow 
to confirm or negate this suggestion. In the positive case, the combination of RV measurements
and astrometry will instantly allow to compute a full set of orbital parameters, and to derive
precise component masses. The knowledge of the orbit will enable the precise reanalysis
of the Hipparcos measurements, resulting in much smaller errors for
the parallax, distance, and distance module. This in return will make the HD\,160934 system 
a valuable calibrator for pre-main-sequence stellar models. \\
The presented data are a good example for the power of the Lucky Imaging technique
as a simple but effective tool for the discovery and follow-up of close binaries. Observations
near the diffraction limit in the visible are possible with a fraction of the instrumental
effort compared to adaptive optics systems or spaceborne observatories.

\begin{acknowledgements}
We are particulary grateful to Sam (Karl) Wagner, Jens Helmling and Uli Thiele
for their help in preparing and commissioning AstraLux as well as all the
technical staff at Calar Alto and MPIA involved in the project.\\
This publication makes use of data products from the Two Micron All Sky Survey, 
which is a joint project of the University of Massachusetts and the Infrared Processing 
and Analysis Center/California Institute of Technology, funded by the National Aeronautics 
and Space Administration and the National Science Foundation.
\end{acknowledgements}

\end{document}